\documentclass[12pt]{article}

\title{The Coupling of Gravity to Spin and Electromagnetism\thanks{This
essay received an ``honorable mention'' in the 1999 Essay Competition of the
Gravity Research Foundation -- Ed.}}
\author{Felix Finster,\ 
Joel Smoller\thanks{Research supported in part by the NSF, Grant No.\ 
DMS-G-9802370.}, and Shing-Tung Yau\thanks{Research supported in part 
by the NSF, Grant No.\ 33-585-7510-2-30.}}
\date{March 1999}

\newcommand{\spc}{\;\;\;\;\;\;\;\;\;\;}

\newcommand{\1}{\mbox{\rm 1 \hspace{-1.05 em} 1}}

\begin{document}
\include{epsf}

\maketitle
\begin{abstract}
The coupled Einstein-Dirac-Maxwell equations are considered for a 
static, spherically symmetric system of two fermions in a singlet 
spinor state. Stable soliton-like solutions are shown to exist, and 
we discuss the regularizing effect of gravity from a Feynman diagram 
point of view.
\end{abstract}

There are some interesting effects that result when one 
couples gravity, as expressed through Einstein's equations, to other 
fundamental force fields. The most striking example of this is the 
discovery by Bartnik and McKinnon (BM) \cite{BM} of
soliton-like solutions for the $SU(2)$ Einstein-Yang/Mills (EYM) 
equations. This came as a surprise because several results for 
related systems lead to the conjecture that such solutions cannot 
exist \cite{De}. For example, neither the vacuum Einstein 
equations, nor the Yang/Mills equations admit non-trivial static, 
globally regular solutions. The existence of the BM solutions, 
however, is due to the coupling of different force fields, whereby 
the attractive gravitational force is balanced by the repulsive 
Yang/Mills force. But this balance is quite delicate, so much so, that 
the EYM solutions are unstable to small perturbations \cite{SZ}.

We consider here a different type of coupling; namely gravity coupled 
both to quantum mechanical particles with spin (Dirac particles) and
to an electromagnetic field. We analyze the resulting 
Einstein-Dirac-Maxwell (EDM) equations for a static, 
spherically symmetric system of two fermions in a singlet spinor state. 
We construct {\em{stable}} soliton-like solutions, and we 
discuss some of their properties for different values of the electromagnetic 
coupling constant. Our aim is to show that
the gravitational interaction has a certain ``regularizing'' effect;
this will be explained from a Feynman diagram point of view.

Before going further, we point out that throughout this essay, we are 
considering only {\em{classical}} fields coupled to the Dirac 
equation; in particular, we consider neither quantized bosonic 
fields, nor second quantization of the Dirac field.

The EDM equations for a system of $n$ Dirac particles
have the following general form,
\begin{equation}
\left. \begin{array}{rcll}
R^i_j \:-\: \frac{1}{2}\:R\: \delta^i_j &=& -8 \pi\: T^i_j&
{\mbox{(Einstein equations)}} \\[.2cm]
(i G^j D_j -m) \:\Psi_a &=& 0 \;\;\;,\;\;\;\;\;\; a=1,\ldots,n \spc
& {\mbox{(Dirac equations)}} \\[.2cm]
\nabla_l F^{kl} &=& 4 \pi \:j^k &{\mbox{(Maxwell equations)}}
\end{array} \right\} \;, \label{EDM}
\end{equation}
where $j^k=e \sum_{a=1}^n \overline{\Psi_1} G^k \Psi_a$
is the Dirac current, and $T^i_j$ is the sum of the energy momentum
tensor of the Dirac particles and of the Maxwell stress-energy tensor.
Here $R^i_j$ is the Ricci tensor, $F_{jk}$ is the electromagnetic (em)
field tensor, and $\Psi_a$ are the 4-component Dirac wave functions
describing particles of mass $m$ and charge $e$.
The Dirac matrices $G^j$ are $(4 \times 4)$ matrices, which are related to
the Lorentzian metric via the anti-commutation relations
$\frac{1}{2} \{G^j(x), G^k(x)\} = g^{jk}(x)\:\1$. $D_j$ is the spin connection;
it is composed of the Christoffel symbols of the metric connection and of the 
electromagnetic potential (see \cite{F}).
The EDM equations only make physical sense if the wave functions $\Psi_a$ are
properly normalized. For this, we choose a space-like hypersurface
${\cal{H}}$ with a future-directed normal vector field $\nu$ and demand that
\[ \int_{\cal{H}} \overline{\Psi_a} G^j \Psi_b \:\nu_j \:d\mu \;=\;
\delta_{ab} \;\;\;,\spc a,b=1,\ldots, n, \]
where $d\mu$ is the invariant measure on ${\cal{H}}$, induced by the Lorentzian
metric.

The Dirac and Maxwell equations are linear in the wave functions and 
in the electromagnetic field, respectively. But the Dirac current and the 
energy-momentum tensor are quadratic in the $\Psi_a$, the 
stress-energy tensor is quadratic in the electromagnetic field, and
the Einstein equations are nonlinear in the metric. Thus the EDM equations
form a highly nonlinear system of differential equations, which
in this generality is far too complicated to analyze.
Therefore we consider a simple
ansatz where the EDM equations reduce to a system of ordinary differential
equations; namely a static, spherically symmetric system of two Dirac
particles in a singlet spinor state. For the wave functions, ``static'' 
means that their time dependence is a plane wave $e^{-i \omega t}$. The
parameter $\omega>0$ is the energy of the Dirac particles.
Using a special rescaling technique \cite{FSY1, FSY2}, we construct numerical
solutions of our EDM system.

\begin{figure}[tb]
	\epsfxsize=13cm
	\centerline{\epsfbox{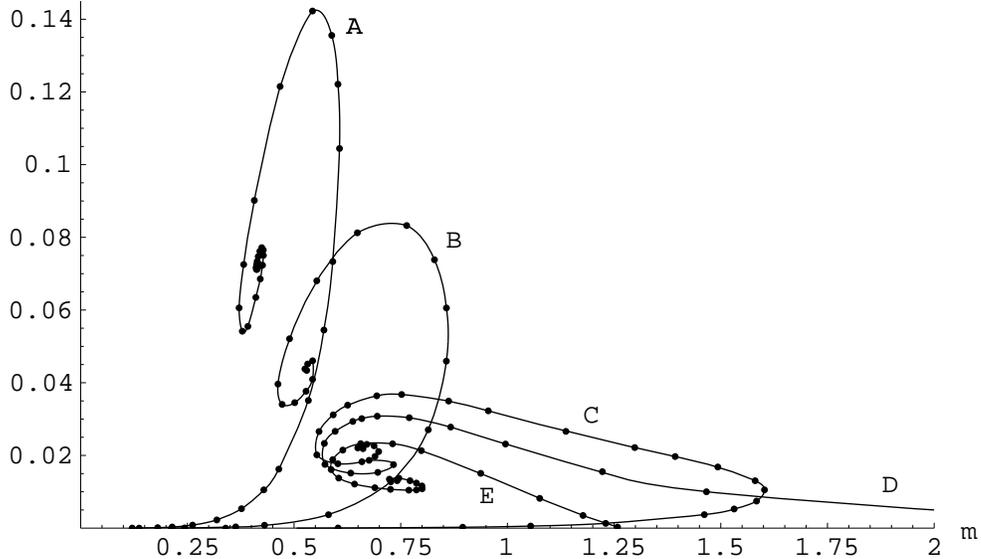}}
	\caption{Binding Energy $m-\omega$ of the Fermions for $(e/m)^2=0$ 
	(A), $0.7162$ (B), $0.9748$ (C), $1$ (D), and $1.0313$ (E).}
	\label{EM_bind}
\end{figure}

We now briefly mention some
general properties of the solutions. All of our solutions are everywhere
regular, asymptotically Minkowskian, and have finite (ADM) mass $\rho$.
The Dirac wave functions decay
exponentially at infinity, so that the Dirac particles are with a high
probability confined to a neighborhood of the origin; this is due to 
the gravitational attraction. For large $r$, our
solutions look similar to a Reissner-Nordstr\"om solution with charge $2e$ and
mass $\rho$.
Our solutions are parametrized by two physical parameters: the rest mass $m$
of the Dirac particles and the relative strength of the electromagnetic and
gravitational coupling $(e/m)^2$. For a fixed value of these parameters, we
find an infinite, discrete number of solutions which
can be regarded as the ground state, the first excited state, etc.
In Figure~\ref{EM_bind}, the mass-energy spectrum for the ground state;
i.e., the binding energy $m-\omega$ versus $m$, is plotted 
for different values of $(e/m)^2$; these curves are spirals.
The binding energy decreases for fixed $m$ as $(e/m)^2$ increases; this means
that the em repulsion weakens the binding.
For $(e/m)>1$, the electromagnetic repulsion is so strong that there 
are no solutions in the Newtonian limit. However, due to the nonlinearity
of Einstein's equations, the EDM equations still admit solutions (spiral E).
By combining linear perturbation theory with more subtle topological 
methods (Conley index, cf.\ \cite{Sm}), the stability of the solutions is analyzed 
\cite{FSY1}. For $(e/m)^2 <1$, we find that all solutions on the ``lower 
branch'' of the spiral (i.e.\ on the curve from $m=0$ up to the maximal 
value of $m$) are stable, whereas all solutions on the ``upper branch'' 
are unstable.

\begin{figure}[tb]
	\centerline{\epsfbox{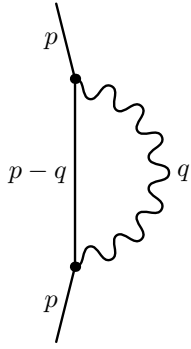}}
	\caption{A simple loop diagram}
	\label{loop}
\end{figure}

Our solutions of the EDM equations have some interesting
features. In this essay, we shall focus on the discussion of what the 
form of the spiral curves in Figure \ref{EM_bind} tells us about the 
nature of the gravitational interaction (see 
\cite{FSY1,FSY2} for the discussion of other effects).
If the EDM equations (\ref{EDM})
are treated in a perturbation expansion (in the spirit of \cite{BD}),
we obtain a description of the interaction with Feynman diagrams.
More precisely, we get all the Feynman diagrams of quantum field theory,
with the only exception of the fermionic loop diagrams; these are 
absent because we do not consider pair creation/annihilation (the number of
Dirac particles is fixed in our ansatz). Thus we can say that, although
we consider only classical fields, many aspects of perturbative quantum
field theory are already incorporated in our EDM equations.
The first question arising from this point of view is why the divergences
of the Feynman diagrams do not show up in our regular, well-behaved
solutions. To analyze this problem in a simple example, we consider
the loop diagram of the virtual emission and absorption of a photon in
Figure \ref{loop}.
If evaluated in Minkowski space, this diagram involves
the divergent integral over the product of two Green's functions,
\begin{equation}
\lim_{\varepsilon \searrow 0} \int d^4q \:\frac{(p^j - q^j) \gamma_j + m}
{(p-q)^2 - m^2 - i \varepsilon p^0} \; \frac{1}{(q^2 - i \varepsilon 
q^0)} \;\;\;. \label{d}
\end{equation}
In our ansatz, we only consider a static electromagnetic potential, 
which means that the momentum $q$ of the electromagnetic Green's function
must be purely spatial, $q=(0, \vec{q})$. Thus the process is described,
instead of (\ref{d}), by the three-dimensional integral
\begin{equation}
\lim_{\varepsilon \searrow 0}
\int d\vec{q} \:\frac{p^0 \gamma_0 + (\vec{p}-\vec{q}) \vec{\gamma} + m}
{(p^0)^2 - (\vec{p}-\vec{q})^2 - m^2 -i \varepsilon p^0}
\; \frac{1}{(-\vec{q}^2 -i \varepsilon q^0)} \;\;\; .
\end{equation}
This integral is finite, as one sees by considering the behavior of the
integral for large $|\vec{q}|^2$, taking into account spherical symmetry.
This illustrates that the reason why our interaction is finite is
because of our restriction to static, spherically symmetric solutions;
note that it is unrelated to the fact that our fields are merely
classical.

We now come to our main point: the convergence of the perturbation
expansion. A perturbation expansion in the gravitational coupling constant
corresponds to a Taylor expansion of our system in the mass parameter 
$m$. We consider as an example a spiral of Figure~\ref{EM_bind} for small 
(fixed) $(e/m)^2$. The just explained finiteness of all Feynman
diagrams becomes manifest by the fact that the spiral is
smooth near the origin. However, since it is not even one-to-one, the 
spiral is {\em{not}} analytic in $m$. This implies that the perturbation
expansion cannot converge absolutely for large values of the coupling constant.
This explicit non-convergence result is a peculiar effect of the
gravitational interaction; e.g. the mass-energy spectrum for the 
Dirac-Maxwell system \cite{L} is analytic. Nevertheless, since the 
spiral curves are bounded, we see that the total
interaction is finite even for strong coupling. This implies that 
there {\em{must be}} cancellations between Feynman diagrams of different order,
such that the perturbation series is finite, although this series does not
converge absolutely. Thus the gravitational interaction of the Dirac
particles is more regular 
than what one would expect from a naive perturbation expansion.
Actually, it has long been conjectured that the ultraviolet
divergences of quantum field theory might disappear
when gravitation is included in a non-perturbative way (see \cite{T} for a
recent formulation of this conjecture).
Our EDM system allows us to verify this conjecture explicitly in a
simple example.
Our results also show that effects like conditional summability of the
perturbation series actually occur for the fundamental physical interactions
with gravity.

\begin{footnotesize}
\hspace*{-.65cm} 
\begin{tabular}{lclcl}
\\
Felix Finster && Joel Smoller && Shing-Tung Yau \\
Max Planck Institute MIS && Mathematics Department && Mathematics 
Department \\
Inselstr.\ 22-26 && The University of Michigan && Harvard University \\
04103 Leipzig, Germany && Ann Arbor, MI 48109, USA
&& Cambridge, MA 02138, USA \\
Felix.Finster@mis.mpg.de && smoller@umich.edu && yau@math.harvard.edu
\end{tabular}
\end{footnotesize}

\end{document}